\documentclass[npg]{copernicus}
\begin{document}
\title{Observational Tests of the Properties of Turbulence in the Very Local Interstellar Medium}
\author[1]{Steven R. Spangler}
\author[1]{Allison H. Savage}
\author[2]{Seth Redfield}
\affil[1]{Department of Physics and Astronomy, University of Iowa, Iowa City, IA 52242}
\affil[2]{Wesleyan University, Astronomy Department \& Van Vleck Observatory, Middletown, CT 06459}
\runningtitle{Turbulence in Very Local ISM}
\runningauthor{Spangler et al}
\correspondence{Steven R. Spangler\\ steven-spangler@uiowa.edu}
\received{}
\pubdiscuss{} %% only important for two-stage journals
\revised{}
\accepted{}
\published{}
\firstpage{1}
\maketitle
\begin{abstract}
The Very Local Interstellar Medium (VLISM) contains clouds which consist of partially-ionized plasma.  These clouds can be effectively diagnosed via high resolution optical and ultraviolet spectroscopy of the absorption lines they form in the spectra of nearby stars.  Among the information provided by these spectroscopic measurements is $\xi$, the root-mean-square velocity fluctuation due to turbulence in these clouds, and $T$, the ion temperature, which may be partially determined by dissipation of turbulence.  We consider whether this turbulence resembles the extensively studied and well-diagnosed turbulence in the solar wind and solar corona. Published observations are used to determine if the velocity fluctuations are primarily transverse to a large-scale magnetic field, whether the temperature perpendicular to the large scale field is larger than that parallel to the field, and whether ions with larger Larmor radii have higher temperatures than smaller gyroradius ions. Our approach is to determine if the spectroscopically-deduced parameters such as $\xi$ and  $T$ depend on direction on the sky.  We also consider the degree to which a single temperature $T$ and turbulence parameter $\xi$ account for the spectral line widths of ions with a wide range of masses.  Although a thorough investigation of the data is underway, a preliminary examination of the published data shows neither evidence for anisotropy of the velocity fluctuations or temperature, nor Larmor radius-dependent heating.  These results indicate differences between solar wind and Local Cloud turbulence.  Possible physical reasons for these differences are discussed.      
\end{abstract}
\introduction
It is arguably the case that the best experimental data for studying magnetohydrodynamic (MHD) turbulence come from spacecraft measurements of turbulent fluctuations in the solar wind. The obvious advantage of  solar wind turbulence is that basic plasma physics measurements of vector magnetic field, plasma flow velocity, density, temperatures, and even electron and ion distribution functions can be measured in situ with spacecraft.  The fluctuations in all of these quantities have been extensively studied, a large literature written, and major conclusions reached. Among the many reviews of the subject is the monograph by \cite{Tu95} and the review article by \cite{Goldstein95}. 

Another nearby plasma with extensive diagnostics (although not, as yet, in situ measurements) is the solar corona.  Our knowledge of the corona and its turbulence results from high spatial resolution images, ultraviolet spectroscopy of numerous transitions, and radio propagation measurements. In addition, a sort of ``ground truth'' for coronal plasma measurements is provided by spacecraft measurements at heliocentric distances of 0.28 to 1 astronomical units, and beyond.  The coronal plasma is convected out into space and becomes the solar wind.   Among the many reviews of the coronal plasma, two which are particularly relevant to the present investigation are \cite{Cranmer02} and \cite{Bird90}.  

Measurements and observations of solar wind and coronal turbulence provide the best data for testing our ideas on the nature of MHD turbulence, and guide our thinking about turbulence in other astrophysical plasmas.  In fact, one could contend that astronomers and astrophysicists, when considering astronomical processes which may be due to turbulence (e.g. support of molecular clouds against their own gravity, moderation of heat conduction in clusters of galaxies, or instabilities in accretion disks) should begin with the assumption that the turbulence of interest resembles the well-observed turbulence in the solar wind.  

In this paper, we discuss the degree to which the turbulence which exists in one well-diagnosed plasma, the partially ionized media which constitute the Local Clouds in the Very Local Interstellar Medium (VLISM), have properties similar to those of turbulence in the solar wind and solar corona. 
\section{Main Characteristics of Solar Wind Turbulence}
A list of the principal results on solar wind turbulence could be long, idiosyncratic of the individual making the compilation, or both.  Four which we consider important and relevant to the present investigation are as follows. 
\begin{enumerate}
\item The fluctuations in magnetic field and plasma flow velocity are primarily perpendicular to the large scale magnetic field. This property also emerges from theories of MHD turbulence, specifically quasi-2D turbulence \citep{Strauss76, Zank92, Spangler99}.  Observational evidence for this property comes from \cite{Bavassano82} and \cite{Klein93}.  A summary and reference to other papers is given in \cite{Tu95}.  The clearest evidence for these anisotropic, transverse fluctuations appears in the high-speed solar wind \citep{Bavassano82, Klein93}.  Under such circumstances, the ``minimum variance direction'' is very close to aligned with the large scale magnetic field, and the magnetic field variance in a direction perpendicular to the field is several times larger than the variance of the component in the direction of the large scale field.  See Figure 2 of \cite{Bavassano82} for an illustration of these results. 

These characteristics are not universal in the solar wind.  As discussed by \cite{Klein93} and \cite{Tu95}, the fluctuations are more isotropic in the slow speed solar wind, and anisotropy and field alignment deteriorate with increasing heliocentric distance.  A discussion of the possible physical mechanisms for this evolution is given by \cite{Klein93}. 
\item The irregularities (fluctuations in any plasma quantity) are stretched out along the large scale field.  This was shown to be the case by \cite{Strauss76} for the case of Tokamaks, and is now appreciated to be a general property of MHD turbulence, at least when the amplitude of the turbulence is modest. Among the best observational illustrations of this effect in an astrophysical plasma are radio propagation measurements of the solar corona, which show density fluctuations (presumably passive tracers of the turbulent velocity fluctuations) drawn out along the coronal magnetic field with an axial ratio than can exceed 10 \citep{Armstrong90, Coles03}.  
\item The interaction of turbulence with ions leads to perpendicular (to the large scale magnetic field) temperatures which are higher than parallel temperatures, or at least larger than would be expected without effective perpendicular heating \citep{Cranmer02, Kasper09}.  
\item Interaction of turbulence with ions leads to temperatures which depend on the ion species; ions with larger Larmor radii have higher temperatures \citep{Cranmer02}.  This is an observational signature of cyclotron resonance heating \citep{Hollweg08}, and is more pronounced in the corona than the solar wind, although it can be found in both media.   
\end{enumerate}
\section{Well-Diagnosed Plasmas Beyond the Solar System}
Much of astronomy and astrophysics consists of the study of plasma media beyond the solar system.  In all cases the quality of the information on the plasma state in an astrophysical medium is inferior to that available for the solar wind at the orbit of the Earth.  However, in some cases, the quality of astronomical observations is sufficient to provide diagnostics of the plasma which would be acceptable by plasma physics laboratory standards.  Two media which are particularly noteworthy in this regard are the Diffuse Ionized Gas (DIG) phase of the interstellar medium, which is believed to fill the largest portion of the volume of the interstellar medium, and the Local Clouds in the interstellar medium, which comprise the closest samples of the interstellar medium. 
\subsection{The Diffuse Ionized Gas (DIG)}
Approximately 25 \% of the interstellar medium is filled with a gas possessing a density of about 0.1 cm$^{-3}$, a temperature of $\simeq$ 8000K, and a magnetic field of 3-4 $\mu$ G. This phase of the interstellar medium is referred to as the Diffuse Ionized Gas, or DIG. There are a number of diagnostics which show the existence of this medium, and provide numbers for its plasma properties.   The most important of these diagnostics has been extremely high spectral resolution Fabry-Perot interferometry of the H$\alpha$ glow from the plasma, done by dedicated instruments at the University of Wisconsin \citep[for a description of the most recent version of the instrument, as well as references to previous literature and results, see ][]{Haffner03}. Radio propagation measurements such as pulsar dispersion measures and Faraday rotation measures of extragalactic radio sources and pulsars also provide information.  A somewhat old but still valuable review of our knowledge of the DIG is given by \cite{Cox87}. 

From high resolution optical spectroscopic and radio propagation measurements, we have been able to determine an impressive number of plasma parameters in the DIG.  These include the electron density, the ionization fractions of hydrogen and helium, the electron temperature, the ion temperature, the strength of the magnetic field as well as some information on its spatial fluctuations, and insight into the heating mechanisms which are operative.  

The  properties of the plasma in the DIG phase of the interstellar medium are listed in Table 1. 
\begin{table*}[t]
\caption{Mean Plasma Parameters in Diffuse Ionized Gas (DIG) Phase of the Interstellar Medium}
\vskip4mm
\begin{tabular}{ll}
\tophline
Plasma Parameter & Value \\
\middlehline
electron density & 0.08 cm$^{-3}$ \\
hydrogen ionization fraction &  $\geq$ 0.90 \\
ion temperature &  8000 K (typical) \\
magnetic field & 3-4 $\mu$ G \\  
\bottomhline
\end{tabular}
\end{table*}

\subsection{Local Clouds in the Very Local Interstellar Medium (VLISM)}
Another example of a well-diagnosed plasma medium beyond the solar system is that of the small clouds in the vicinity of the Sun.  These clouds are embedded within the Local Cavity of the interstellar medium, and lie within a few parsecs of the Sun.  The Local Cavity is a region of low gas density that extends for approximately 100 parsecs from the Sun. ``Tomographic'' representations of this cavity, determined with similar observational techniques and undertaken by the same group, are presented in \cite{Lallement03} and \cite{Welsh10}. 

The presence of clouds in the Local Cavity is indicated by non-stellar absorption lines in the spectra of nearby stars.  Reviews of the properties of these clouds are given by \cite{Frisch00} and \cite{Redfield09}. 

The typical plasma properties of these clouds are given in Table 1. 
\begin{table*}[t]
\caption{Mean Plasma Parameters of Local Interstellar Clouds}
\vskip4mm
\begin{tabular}{ll}
\tophline
Plasma Parameter & Value \\
\middlehline
electron density & 0.11 cm$^{-3}$ \\
neutral density &  0.1 cm$^{-3}$  \\
temperature &  4000-8000 K (typical) \\
magnetic field & 3-4 $\mu$ G (assumed)\\  
\bottomhline
\end{tabular}
\end{table*}
\cite{Redfield08a}  have discovered 15 distinct clouds within 15 parsecs of the Sun. 
Comparison of Tables 1 and 2 indicates that, at a local level, the DIG and Local Clouds are somewhat similar.  A possibly significant difference is in the degree of ionization.  In the DIG, the hydrogen is probably completely ionized, and neutral atoms are restricted to partially ionized helium.  In the Local Clouds, the hydrogen is only about 50 \% ionized.  As a result, ion-neutral collisional processes will be much more important in the Local Clouds.  For the remainder of this paper, we will be concerned with plasma processes in the Local Clouds.  We will be particularly interested in properties of turbulence in the Local Clouds, and the degree to which this turbulence resembles, or differs from, that in the solar wind and solar corona.  
\section{Turbulence Diagnostics from High Resolution Spectroscopy}
It is at first remarkable that any information on turbulence in the Local Clouds can be obtained from measurements of absorption lines in the spectra of background stars.  The way in which this can be done is as follows.  Spectroscopic measurements are made of the Doppler shift, column density, and line width of the absorption lines.  \cite{Redfield04} report measurements of absorption lines for as many as 8 transitions of neutral atoms and ions.  The different atoms or ions have different values of the atomic mass $m$.

With these measurements, they fit a model for the line width as a function of atomic mass $m$ of the following form. 
\begin{equation}
b^2 = \frac{2 k_B T}{m} + \xi^2
\end{equation}
where $b$ is the measured line width of a transition and absorption component\footnote{Many lines of sight to nearby stars show different absorption components at different Doppler shifts.  The Doppler components are interpreted as absorption in distinct Local Clouds.}, $T$ is the ionic or atomic temperature, and $\xi$ is the nonthermal broadening parameter.  Equation (1) states that the measured line width is a quadratic sum of two Doppler shifts, the first due to thermal motion of the atoms or ions, and the second due to atomic and ionic motions not attributable to thermal effects, and which are interpreted as turbulent motions in which all species take part.  
\cite{Redfield04} show a comparison of the model fit to the data for many absorption components and lines of sight in Figure 1 of their paper.  The quality of the fits is impressively good. 

To anticipate one of the main points of this paper, a coronal astronomer or solar wind physicist would immediately take issue with Equation (1), noting point (4) above that in solar system media, different atoms and ions have different temperatures.  In Section 5.3 we will investigate the degree to which a single, common temperature characterizes the Local Clouds.  
\section{Does Turbulence in the Local Clouds Resemble that in the Solar Wind?}
From the data in Table 1 of \cite{Redfield04}, we have measurements of $T$ and $\xi$ for 53 absorption components along 32 lines of sight to nearby stars distributed around the sky.  In the remainder of this section, we consider what these measurements can tell us about turbulence in the Local Clouds, and the degree to which it resembles heliospheric turbulence. 
\subsection{A Test for Transverse Velocity Fluctuations}
We first consider whether the turbulence in the Local Clouds possesses transverse velocity fluctuations, in which the fluid motion is perpendicular to the large scale magnetic field.  The way in which this might reveal itself in the data of \cite{Redfield04} is shown in Figure 1.  
\begin{figure}[h]
\vspace*{2mm}
\includegraphics[width=8.0cm]{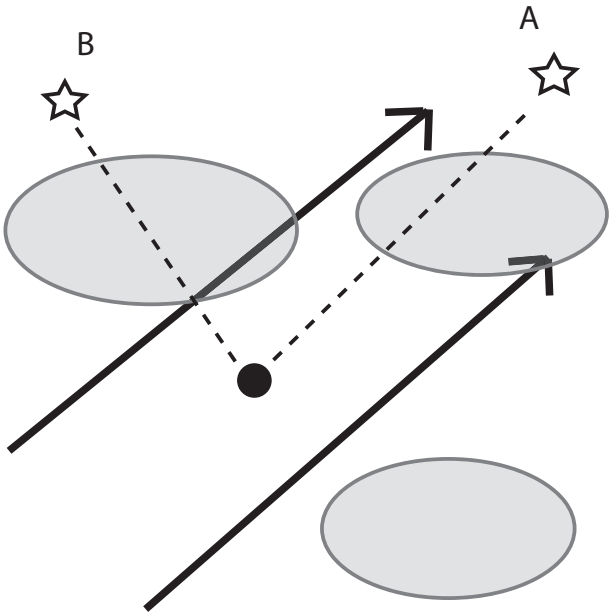}
\caption{Cartoon illustrating model of the plasma in the Very Local Interstellar Medium.  We envision relatively dense, partially-ionized clouds (indicated by the gray-shaded regions) embedded in the lower density Local Cavity. The cavity and its clouds possess a magnetic field which is relatively uniform in magnitude and direction (heavy lines) on a scale which includes all clouds found to date.  Turbulence in the clouds is magnetohydrodynamic and Alfv\'{e}nic in nature, with velocity and magnetic fluctuations primarily transverse to this large scale field. Lines of sight along the field (to star A) should show relatively small turbulent broadening, whereas lines of sight across the field (to star B) show should large values of turbulent broadening.}
\end{figure}
We assume that one can define a large scale magnetic field which is constant in direction over spatial scales which include all of the Local Clouds. This direction is defined by the unit vector $\hat{b} \equiv \frac{\vec{B}}{B}$. The scale of magnetic field uniformity should be of the order 30 parsecs.  For some lines of sight, we are looking along this large scale field.  In this case, the turbulent motions will be primarily perpendicular to the line of sight, and little turbulent line broadening should be observed ($\xi$ small).  Other lines of sight will be perpendicular to the large scale field, and the turbulent velocity fluctuations will be along the line of sight.  In this case, $\xi$ will be large.  

We now present a quantitative model for the dependence of the turbulent line width $\xi$ on direction on the sky. The turbulence model we want to test is one in which the fluctuations in the plane perpendicular to $\hat{b}$ are larger than those in the direction of $\hat{b}$. A probability distribution function which is simple in mathematical form and describes this is 
\begin{equation}
f(\vec{v}) = \left( \frac{1}{(2 \pi)^{3/2} V_{\parallel} V_{\perp}^2} \right) \exp (-\frac{v_z^2}{2 V_{\parallel}^2}) \exp (-\frac{v_x^2 + v_y^2}{2 V_{\perp}^2}) 
\end{equation}
with $V_{\perp} > V_{\parallel}$ by assumption. The z axis in Equation (2) is in the direction of $\hat{b}$. The distribution function Equation (2)  satisfies the normalization requirement that $\int d^3 v f(\vec{v}) = 1$. 

The observed spectral line width is proportional to the rms fluctuation in the component of the velocity along the line of sight, which is in the direction of the unit vector $\hat{l}$.  The unit vector $\hat{l}$ points in the direction of galactic longitude and latitude $(l,b)$.  In what follows, we assume that the observed turbulent width $\xi$ can be expressed as
\begin{equation}
\xi^2 = <v_L^2> = <v_L^2(v_x,v_y,v_z)> = \int d^3v v_L^2 (v_x,v_y,v_z)
\end{equation}
By expressing $\xi^2$ as an expectation value, we assume that the line of sight integration through the cloud samples a large number of independent eddies in the cloud and thus satisfies the ergodic theorem.  

It is necessary to express the line-of-sight component of the velocity in terms of the magnetic-field-oriented system of coordinates, $v_L(v_x,v_y,v_z)$.  This is done by a set of Euler angle transformations which transform from the direction of the large scale field $\hat{b}$ to that of the line of sight $\hat{l}$.  The details of this transformation are given in Spangler, Savage, and Redfield (2010, in preparation).  The result of this calculation is the following expression for  $<v_L^2(v_x,v_y,v_z)>$, or equivalently, the observed turbulent line width $\xi$. 

\begin{equation}
\frac{<v_L^2>}{V_{\perp}^2} = \frac{\xi^2}{V_{\perp}^2} = 1 - \epsilon (\sin b \sin \beta + \cos \Delta l \cos b \cos \beta )^2
\end{equation}
Variables in Equation (4) which have not been previously defined are $(\lambda, \beta)$, the Galactic latitude and longitude of the direction of the magnetic field,  $\Delta l \equiv \lambda - l$, and the anisotropy parameter $\epsilon \equiv 1 - \frac{V_{\parallel}^2}{V_{\perp}^2}$. Isotropy of turbulence corresponds to $\epsilon = 0$, while $\epsilon =1$ means the velocity fluctuations lie in a plane perpendicular to $\hat{b}$. 

Equation (4) is convenient for describing how $\xi$ depends on position on the sky (provided $(\lambda, \beta)$ were known; see below), but a more instructive expression is in terms of the angle $A$ between the line of sight and the large scale magnetic field, $\cos A \equiv \hat{l} \cdot \hat{b}$,
\begin{equation}
\frac{<v_L^2>}{V_{\perp}^2} = \frac{\xi^2}{V_{\perp}^2} = 1 - \epsilon \cos^2A
\end{equation}

The form of $\frac{\xi^2}{V_{\perp}^2}$ as function of $A$ for various values of $\epsilon$ is shown in Figure 2. 
 
\begin{figure}[h]
\vspace*{2mm}
\includegraphics[width=8.0cm]{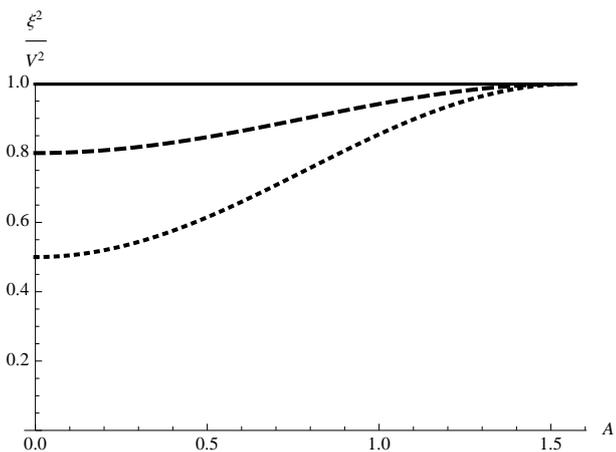}
\caption{Expected dependence of the square turbulent line width (normalized by the turbulent velocity amplitude) $\frac{\xi^2}{V_{\perp}^2}$ as a function of direction on the sky.  The abscissa is the angle $A$ (in radians) between the line of sight and the direction of the local interstellar magnetic field.  The three curves correspond to values of $\epsilon=0$ (solid line), 0.2 (dashed line), and 0.5 (dotted line).}
\end{figure}

An analysis based on Equations (4) and (5) is complicated by the fact that we do not definitely know the direction of the large scale field, $\hat{b}$, although there are independent estimates for this direction \citep{Gurnett06, Opher09}.  For the present, we can ask if there is {\em any} direction in the sky for which the turbulent velocity broadening data adhere to the form of Equation (5).  We are presently carrying out an analysis of the data given in \cite{Redfield04} to answer this question. However, a preliminary conclusion is provided by Figure 9 of \cite{Redfield04}, which plots the derived values of $\xi$ on a map of the sky. This figure shows no obvious way in which the $\xi$ data could be organized as indicated by Equation (5).  That is, there are no clear poles characterized by low values of $\xi$, with a corresponding band across the sky with larger values of $\xi$.  Instead, there appear to be random variations of $\xi$ values about a mean value.  

Although a more definitive result will be available after we complete our quantitative analysis of the data in \cite{Redfield04}, at the present we can say that there is no strong evidence for anisotropy of turbulent velocity fluctuations in the Local Clouds, as would occur if the turbulence is transverse in the sense of solar wind turbulence.  
\subsection{A Test for Enhanced Perpendicular Ion Heating}
We can also examine whether the thermal temperature $T$ depends on direction on the sky, as would be expected if the temperature perpendicular to the large scale magnetic field were larger than that parallel to the field.  Formulas similar to those presented in Section 5.1 should hold in this case as well; in both cases the quantity considered ($<v_L^2>$ or T) corresponds to the second moment of a velocity distribution function.  In the case of the turbulent velocity fluctuations, the expression Equation (2) was adopted as a model.  In the case of the temperature, a Maxwellian should be an exact expression.  

We then ask if the ion temperature measurements show a dependence on direction on the sky similar to Equation (5). We are presently carrying out a quantitative analysis of this point as well.  However, a preliminary and qualitative answer is provided by the results in \cite{Redfield04}.  Figure 8 of \cite{Redfield04} shows the retrieved ion temperature $T$ as a function of position on the sky for the 53 absorption components.  The values appear to show random fluctuations about a mean value.  The agent responsible for these variations from one line of sight to another is unknown, but the variations do not seem to be of the form in Equation (5).  We conclude that there is no evidence for departure of the ratio $\frac{T_{\perp}}{T_{\parallel}}$ from unity in these clouds.  
\subsection{A Test for Mass-Proportional Ion Heating}
The final issue we wish to consider is whether the plasma medium of the Local Clouds shows a dependence of ion temperature on ion mass, as is strongly the case in the corona, and to a lesser but significant degree in the solar wind.  As noted in Section 2, this is an indicator of ion cyclotron resonance effects in the interaction between turbulence and the particles which make up the plasma.  

The results of  \cite{Redfield04} are consistent with a single temperature for all species, and show no evidence for the presence of ion cyclotron resonance heating.  The basis of this statement is the good agreement of the line width data for as many as 8 transitions with the formula in Equation (1).  This expression features a single temperature for all species.  The fits of Equation (1) to the spectral line width data are satisfactory in the sense of the value of the reduced $\chi^2$ statistic of the fit to the data.  In this case, there is no basis or justification for considering a more complex model for the ion temperature, and thus line width, as a function of ion mass.  If sufficiently strong, ion-specific heating were occurring, it would be impossible to obtain a satisfactory (in the sense of the reduced $\chi^2$ statistic) fit of Equation (1) to the data.  The interested reader should examine the many such fits in Figure 1 of \cite{Redfield04}; it is striking how well Equation (1) fits the data. 

We are presently in the process of testing the degree to which a departure from Equation (1) is allowed by the data.  As a simple model which incorporates ion-mass-specific heating, we are using the expression
\begin{equation}
b^2 = \frac{2 k_B T_0}{m} \left( \frac{m}{m_0}\right)^d + \xi^2
\end{equation}
This model has three parameters ($T$, $\xi$, and $d$) instead of the two parameters of Equation (1).  The parameter which describes ion-mass-proportional heating is $d$; it describes the way the ion temperature could depend directly on the ion Larmor radius, and thus the ion mass.  The approach used in our analysis is to undertake a set of fits of Equation (6) to the data presented in \cite{Redfield04}, and determine the maximum value of $d$ which allows an acceptable value of the reduced chi-square $\chi^2_{\nu}$.  This work is in progress. 

As a final point, there is a result emergent from the data and analysis of  \cite{Redfield04} which is strange to a plasma physicist.  The consistency of the same temperature $T$ for several  transitions is not only the case for ions of different masses (and thus Larmor radii), but also for both ions and neutrals.  The lines used by \cite{Redfield04} include those from neutral atoms as well as ions.  This indicates that plasma physics processes such as plasma wave-particle interactions, which consist of the response of charged particles to the electric and magnetic fields of plasma waves and turbulence, and which can act only on ions and electrons, are unimportant in the Local Clouds. Alternatively, these processes could be present, but are hidden by other, competing processes such as ion-neutral collisions, which rapidly redistribute energy injected in one ion species to both ion and neutral species.  
\section{Discussion}
Given the assumptions and models assumed in our analysis, we conclude that the turbulence in the partially-ionized plasma of the Local Clouds differs in important respects from that in the solar corona and solar wind.  The discussion above would indicate that this turbulence is not anisotropic with respect to the large scale interstellar magnetic field, but instead is isotropic,  does not show a temperature anisotropy in which $T_{\perp} > T_{\parallel}$, which is an indication of ion-cyclotron resonance in the interaction of turbulence with charged particles, and does not show a dependence of temperature on ion mass, which is also a characteristic of ion cyclotron resonance effects.  Taken at face value, these results point toward diversity in the properties of interstellar turbulence, and suggest that turbulence properties that are sometimes assumed universal because of their prominence in solar wind turbulence might be special cases.     

However, in the remainder of this paper, we wish to discuss some qualifications and reservations to the previous statements.  It is possible that the properties of solar wind turbulence discussed in Section 2 are indeed present in the Local Clouds at a local level, but are disguised in the path-integrated measurements of \cite{Redfield04}.  

The search for anisotropy of the turbulence, and for a high perpendicular-to-parallel temperature ratio, were both predicated  on an ordered, unidirectional magnetic field over the spatial volume occupied by the clouds, as illustrated in Figure 1.  This assumption does not preclude the possibility that the interstellar magnetic field is turbulent.  The local field includes both a true, galactic-scale component as well as a turbulent component.  However, we assume that the vector sum of these two components remains approximately unidirectional on a scale larger than the distribution of clouds.  If this assumption is strongly violated, so that the field is randomized on a length scale much smaller than the diameter of the system of clouds (approximately 30 parsecs), the observational results described in Sections 5.1 and 5.2 would be obtained even if the local turbulent motions are transverse to the local magnetic field, and if locally $T_{\perp} > T_{\parallel}$.  

The resolution of this question depends on the value of the outer scale of turbulence in the VLISM.  If the outer scale is smaller than about 30 parsecs, the field would have different directions in the different clouds.  There are estimates of the outer scale in the interstellar medium, but this quantity is not known with high confidence.  \cite{Minter96} reported evidence for  two outer scales in the DIG,  an outer scale to a fully three dimensional turbulence with a value of about 4 parsecs, and a larger scale of about 100 parsecs for larger, but 2-dimensional eddies. 
Since most of the power in the magnetic field fluctuations is on the larger, $\sim 100$ parsec scales, the \cite{Minter96} result would qualitatively support the magnetic field model we have adopted here.  However, a quantitative modeling of the effect of averaging over the random orientations permitted by  \cite{Minter96} should be done to fully address this point.  

The scales of turbulent fluctuations in the interstellar magnetic field have been investigated more recently by Marijke Haverkorn and colleagues, again using the technique of Faraday rotation \citep{Haverkorn04a,Haverkorn04b,Haverkorn06}. Haverkorn and her group have utilized both measurements of polarization of the galactic nonthermal radiation \citep{Haverkorn04a,Haverkorn04b} as well as Faraday rotation of extragalactic sources \citep{Haverkorn06}. \cite{Haverkorn06} report a difference in the power spectrum of magnetic fluctuations between spiral arms and interarm regions of the Milky Way.

Their estimates of the outer scale in the spiral arms (presumably relevant to the Very Local Interstellar Medium) range from 2 - 17 parsecs; Haverkorn et al do not distinguish between fully three dimensional turbulence and possible composite turbulence consisting of fully three dimensional eddies on small to intermediate scales, and two dimensional, planar turbulence on large scales, terminating in a global outer scale.  If the global outer scale in interstellar turbulence is as small as the 2 - 17 parsec range suggested by Haverkorn and colleagues, it seems probable that the interstellar magnetic field would be sufficiently randomized across the Local Clouds to invalidate our assumption of constant magnetic field direction $\hat{b}$, and eliminate the dependence of the turbulent width $\xi$ on position on the sky given by Equation (5), even if locally the turbulence is transverse to the magnetic field. The same conclusion would apply to the dependence of the ion temperature $T$ on sky position. 

An obvious point, which is nonetheless worth recalling, is that the results of \cite{Minter96}, \cite{Haverkorn04a,Haverkorn04b} and \cite{Haverkorn06} are averages over large volumes of the interstellar medium.  There is no guarantee that those turbulence characteristics apply to the Local Clouds or the VLISM.   

Small scale (outer scale $\leq 15$ parsecs) randomization of the local interstellar magnetic field would explain the absence of the coronal/solar wind properties discussed in Sections 5.1 and 5.2 above.  However, it would not eliminate the observational evidence for ion-mass-dependent temperatures, if such were present locally.  While this question will be addressed more fully in our report in progress, it seems highly plausible that this result is due to the greater collisionality of the Local Clouds relative to the solar wind, and even more, the corona.   As noted above, the hydrogen in the Local Clouds is apparently about 50 \% neutral, meaning ion-neutral collisions between protons and hydrogen atoms may be an important process in distributing energy among species, and eliminating phase space features such as temperature anisotropies. Nonetheless, we consider our results interesting, because the cyclotron periods for all ions of interest are much shorter than the ion-neutral collision timescale in the Local Clouds, so the results of Section 5.3 seem to exclude the possible presence of highly vigorous plasma kinetic processes operating on ion Larmor radius scales.  

This area of investigation will benefit from additional research, made possible by additional, similar quality data on more lines of sight.  One of us (SR) has a program in progress with the repaired STIS (Space Telescope Imaging Spectrograph) spectrograph on the Hubble Space Telescope. These observations will provide FeII and MgII line width measurements (crucial for separating the thermal and turbulent contributions to the line widths) on lines of sight presently lacking such data.  The new data should expand the number of lines of sight probed through the Local Clouds, and facilitate progress on understanding the turbulence in these clouds.      

\conclusions

\begin{enumerate}
\item Multi-transition measurements of absorption line widths due to the Local Clouds provide information about the properties of turbulence in those clouds, and permit inferences about the similarities of this turbulence to that in the solar wind.  
\item We assume that the very local interstellar magnetic field is uniform in direction over a scale large enough to include the 15 clouds discovered by \cite{Redfield08a}.  If this is the case, an anisotropy in the velocity fluctuations, in the sense of being perpendicular to this magnetic field, should map into dependence of the turbulent line width on position on the sky.  
\item In similar fashion, if there is an anisotropy in the ion temperature, such that the perpendicular temperature exceeds the parallel temperature, the inferred thermal broadening of these absorption lines should depend on sky position. 
\item Neither of the aforementioned two properties seem to be present in a casual examination of published data, although a more detailed quantitative analysis is warranted and is in progress.  
\item Finally, the data for the Local Clouds can be examined for a dependence of the ion temperature on ion mass, such as is commonly seen in the solar corona and solar wind. This property appears to be absent for the Local Clouds. 
\item We conclude that the turbulence in the Local Clouds may differ in important ways from that in the solar corona and the solar wind.  The existence of a relatively small (less than or of the order of a few parsecs) outer scale in the Local Cloud turbulence could essentially ``disguise'' solar wind turbulence properties in the data set we have used.  If the outer scale is comparable to or larger than several tens of parsecs, there appears to be a genuine difference in the degree of anisotropy.  The absence of a mass dependence to the ion temperature may be due to the higher collisionality of the Local Clouds relative to heliospheric plasmas. 
\end{enumerate}
\begin{acknowledgements}
This work was supported at the University of Iowa by grants ATM09-56901 and AST09-07911 from the National Science Foundation. 
\end{acknowledgements}


\begin{thebibliography}{}
\bibitem[Armstrong et al (1990)]{Armstrong90} Armstrong, J.W., Coles, W.A., Kojima, M., and Rickett, B.J.: Observations of field-aligned density fluctuations in the inner solar wind, Astrophys. J.~358, 685, 1990 
\bibitem[Bavassano et al (1982)]{Bavassano82} Bavassano, B., Dobrowolny, M., Fanfoni, G., Mariani, F., and Ness, N.F.: Statistical properties of MHD fluctuations associated with high-speed streams from Helios-2 observations, Sol. Phys.~78, 373, 1982 
\bibitem[Bird and Edenhofer (1990)]{Bird90} Bird, M.K. and Edenhofer, P.: Remote sensing observations of the solar corona, in {\em Physics of the Inner Heliosphere}, R. Schwenn and E. Marsch (ed), (Springer-Verlag:Berlin), p13, 1990
\bibitem[Coles et al (2003)]{Coles03} Coles, W.A., Rao, A.P., and Ananthakrishnan, S.: Observations of anisotropic compressive turbulence near the Sun, in {\em Proceedings of the Tenth International Solar Wind Conference}, American Institute of Physics Conference Proceedings~679, 363, 2003 
\bibitem[Cox and Reynolds (1987)]{Cox87} Cox, D.P. and Reynolds, R.J.: The local interstellar medium,  Annu. Rev. Astr. Ap.~25, 303, 1987
\bibitem[Cranmer (2002)]{Cranmer02} Cranmer, S.R.: Coronal holes and the high speed solar wind, Space Sci. Rev.~101, 229, 2002 
\bibitem[Frisch (2000)]{Frisch00} Frisch, P.C.: The galactic environment of the Sun, Am. Sci.~88, 52, 2000 
\bibitem[Goldstein et al (1995)]{Goldstein95} Goldstein, M.L., Robert, D.A., and Matthaeus, W.H.: Magnetohydrodynamic turbulence in the solar wind, Annu. Rev. Astr. Ap.~33, 283, 1995
\bibitem[Gurnett et al (2006)]{Gurnett06} Gurnett, D.A., Kurth, W.S., Cairns, I.H., and Mitchell, J.: The local interstellar magnetic field direction from direction-finding measurements of the heliospheric 2-3 kHz radio emissions, in {\em Physics of the Inner Heliosheath}, American Institute of Physics Conference Proceedings \# 858, p129, 2006 
\bibitem[Haffner et al (2003)]{Haffner03} Haffner, L.M., Reynolds, R.J., Tufte, S.L., Madsen, G.J., Jaehnig, K.P., and Percival, J.W.: The Wisconsin H$\alpha$ Mapper northern sky survey, Astrophys. J. Supp.~149, 405, 2003
\bibitem[Haverkorn et al (2004a)]{Haverkorn04a} Haverkorn, M., Katgert, P., and deBruyn, A.G.: Properties of the warm, magnetized ISM as inferred from WSRT polarimetric imaging, Astron. \& Astrophys.~427, 169, 2004
\bibitem[Haverkorn et al (2004b)]{Haverkorn04b} Haverkorn, M., Gaensler, B.M., McClure-Griffiths, N.M., Dickey, J.M., and Green, A.J.: Magnetic fields and ionized gas in the inner galaxy: an outer scale for turbulence and the possible role of HII regions, Astrophys. J.~609, 776, 2004
\bibitem[Haverkorn et al (2006)]{Haverkorn06} Haverkorn, M., Gaensler, B.M., Brown, J.C., McClure-Griffiths, N.M., Dickey, J.M., and Green, A.J.: Enhanced small-scale Faraday rotation in the galactic spiral arms, Astrophys. J.~637, L33, 2006
 \bibitem[Hollweg (2008)]{Hollweg08} Hollweg, J.V.: The solar wind: our current understanding and how we got here, J. Astrophys. Astr.~29, 217, 2008
\bibitem[Kasper et al (2009)]{Kasper09} Kasper, J.C., Maruca, B.A., and Bale, S.D.: An association between anisotropic plasma heating and instabilities in the solar wind, arXiv: 0911.2715, 2009 
\bibitem[Klein et al (1993)]{Klein93} Klein, L., Bruno, R., Bavassano, B., and Rosenbauer, H.: Anisotropy and minimum variance of magnetohydrodynamic fluctuations in the inner heliosphere, J. Geophys. Res.~98, 17461, 1993 
\bibitem[Lallement (2003)]{Lallement03} Lallement, R.,  Welsh, B.Y., Vergely, J.L., Crifo, F., and Sfeir, D.: 3D mapping of the dense interstellar gas around the Local Bubble, Astron. \& Astrophys..~411, 447, 2003
\bibitem[Minter and Spangler (1996)]{Minter96} Minter, A.H. and Spangler, S.R.: Observation of turbulent fluctuations in the interstellar plasma density and magnetic field on spatial scales of 0.01 to 100 parsecs, Astrophys. J.~458, 194, 1996
\bibitem[Opher et al (2009)]{Opher09} Opher, M. et al: A strong, highly-tilted magnetic field near the solar system, Nature~462, 1036, 2009 
\bibitem[Redfield and Linsky (2004)]{Redfield04} Redfield, S. and Linsky, J.L. 2004: The structure of the local interstellar medium III: temperature and turbulence, Astrophys. J.~613, 1004, 2004
\bibitem[Redfield and Linsky (2008a)]{Redfield08a} Redfield, S. and Linsky, J.L.: The structure of the local interestellar medium IV: dynamics, morphology, physical properties, and implications of cloud-cloud interactions, Astrophys. J.~673, 283, 2008
\bibitem[Redfield and Linsky (2008b)]{Redfield08b} Redfield, S. and Linsky, J.L.: The structure of the local interstellar medium V: electron densities, Astrophys. J.~683, 207, 2008
\bibitem[Redfield (2009)]{Redfield09} Redfield, S.: Physical properties of the local interstellar medium, Space Sci. Rev.~143, 323, 2009
\bibitem[Spangler (1999)]{Spangler99} Spangler, S.R.: Two dimensional magnetohydrodynamics and interstellar plasma turbulence, Astrophys. J.~522,879, 1999
\bibitem[Strauss (1976)]{Strauss76} Strauss, H.R.: Nonlinear, three-dimensional magnetohydrodynamics of noncircular tokamaks, Phys. Fl.~19, 134, 1976 
\bibitem[Tu and Marsch (1995)]{Tu95} Tu, C.Y. and Marsch, E.: MHD structures, waves, and turbulence in the solar wind, Space Sci. Rev.~73, 1, 1995
\bibitem[Zank and Matthaeus (1992)]{Zank92} Zank, G.P. and Matthaeus, W.H.: The equations of reduced magnetohydrodynamics, J. Plasma Phys.~48, 85, 1992
\bibitem[Welsh et al (2010)]{Welsh10} Welsh, B.Y., Lallement, R.,   Vergely, J.L., and Raimond, S.: New 3D gas density maps of NaI and CaII intersetellar absorption within 300 pc, Astron. \& Astrophys..~510, A54, 2010
\end{thebibliography}
\end{document}